\documentclass[prd,showpacs,preprintnumbers,amsmath,amssymb,twocolumn]{revtex4-2}
\usepackage{graphicx} 
\usepackage{amsthm,amsmath,amssymb}
\usepackage{mathrsfs}
\usepackage{color}
\usepackage{appendix}
\usepackage{hyperref}
\date{\today}
\begin{document}
\title{Probing Dark Photons through Gravitational Decoupling of Mass-State Oscillations in Interstellar Media}

\author{Bo Zhang}
\affiliation{Department of Physics, Anhui Normal University, Wuhu, Anhui 241002, China} 
\author{Cui-Bai Luo}
\email{cuibailuo@ahnu.edu.cn}
\affiliation{Department of Physics, Anhui Normal University, Wuhu, Anhui 241002, China}

\begin{abstract}
    We propose a novel mechanism for photon-dark photon mass state oscillations mediated by gravitational separation during propagation through the interstellar medium. This phenomenon establishes a new avenue for the detection of dark matter. By analyzing gravitational lensing data from quasars, we investigate the sensitivity of this approach to dark photons. Our analysis demonstrates constraints of$\epsilon<10^-2$ in the dark photon mass range of $10^{-14}eV$. Furthermore, we propose potential applications of this mechanism to astrophysical systems with strong gravitational fields, such as neutron stars and black hole accretion disks.
\end{abstract}
\maketitle

\section{Introduction}

The Standard Model (SM) has achieved remarkable success over the past decades, yet persistent observational anomalies—including cosmic-ray excesses like the AMS-02 positron anomaly\citep{feng2014ams}-strongly motivate extensions beyond its framework. To address these discrepancies, numerous beyond-the-Standard-Model (BSM) theories have been proposed. The minimal extension involves the introduction of an additional $U(1)_h$ gauge symmetry\citep{holdom1986two}, which may emerge as a low-energy effective theory of ultraviolet-complete models \citep{okun1982limits,redondo2008helioscope,barducci2021dark}. This symmetry naturally predicts a new gauge boson, termed the dark photon $X_{\mu}$, that kinetically mixes with the Standard Model photon $X_{\mu}$. The corresponding extended Lagrangian is\citep{jaeckel2008signatures}:
\begin{align}
    L_{K}=&-\frac{1}{4}F^{\mu\nu}F_{\mu\nu}-\frac{1}{4}X^{\mu \nu}X_{\mu \nu}+\frac{\sin \chi_0}{2}X^{\mu \nu}F_{\mu \nu} \nonumber \\
          &+\frac{m_\chi^2}{2}cos^2\chi_0 X_\mu X^{\mu}-e A_\mu J^{\mu}_{\text{em}},
    \label{eq:lagrangian_k}
\end{align}
where $F^{\mu\nu}=\partial^\mu A^\nu-\partial^\nu A^\mu$ and $X^{\mu\nu}=\partial^\mu X^\nu-\partial^\nu X^\mu$ denote the field strength tensors of the photon and dark photon fields, respectively.
Here, $m_\chi$ represents the dark photon mass, $J_{em}^{\mu}$ is the electromagnetic current, and $\chi_0$ is the mixing angle.

During propagation through a medium, the photon-dark photon interaction states undergo oscillations. For transversely polarized photons, the oscillation probability is given by\citep{danilov2019constraints,an2013new}:
\begin{equation}
    P=\varepsilon^2 \frac{m_\chi^4}{|m_\gamma^2-m_\chi^2|^2},
    \label{eq:Interation_osc_proba}
\end{equation}
where $m_\gamma$ denotes the effective photon mass within the medium, and $\varepsilon$ represents the kinetic-mixing parameter, which associates with $\chi_0$. In the sub-MeV mass regime $(m_\chi<1MeV)$, this oscillation mechanism underpins key dark photon detection strategies, including cosmic microwave background (CMB) analyses\citep{mirizzi2009microwave,mcdermott2020cosmological,caputo2020dark,garcia2020effective}, "light-shining-through-wall" experiments (LSW) \citep{ahlers2008laser}, helioscopes experiments\citep{redondo2008helioscope} , and direct detection efforts \citep{redondo2013solar,an2015direct}. Equation (\ref{eq:Interation_osc_proba}) reveals that when $m_\chi \ll m_\gamma$, the oscillation probability becomes suppressed as $P\propto (m_\chi/m_\gamma)^4$.Consequently, the sensitivity of these experiments is fundamentally limited by the medium’s effective photon mass $m_\gamma$.

In this work, we propose a novel mechanism leveraging gravitationally induced decoupling of photon-dark photon mass eigenstates during propagation. We derive the oscillation probability between these mass states in gravitational fields, which could addresses the medium-induced suppression of oscillations at ultralow dark photon masses, thereby offering a new paradigm for dark photon detection. By modeling gravitational dispersion effects in gravitationally lensed quasar systems, we demonstrate that photon-to-dark-photon oscillations reduce the apparent luminosity of these astrophysical sources. Our calculations indicate that this approach achieves sensitivity to the kinetic-mixing parameter $\varepsilon < 10^{-2}$ for dark photon masses near $10^{-14}eV$. 

In Section \ref{section_2}, we derive the oscillation probability for photon-dark photon mass eigenstates propagating through a medium in gravitational fields. Section \ref{section_3} investigates the suppression of apparent flux in gravitationally lensed quasars induced by the oscillations mechanism. Section \ref{section_4} quantifies the sensitivity of this method, using characteristic scales. Finally, we conclude by summarizing our results and proposing extensions of this framework to probe the oscillation in strong gravitational environments, such as neutron star and black hole.


\section{The Oscillation Model}
\label{section_2}
The dark-photon Lagrangian given in Equation (\ref{eq:lagrangian_k}) contains the kinetic mixing term $X^{\mu \nu}F_{\mu \nu}$. Through the field redefinition ${A,X}\to{A_R,S}$, the mixing term can be diagonalized:
\begin{equation}
    \begin{aligned}
        L_I &= -\frac{1}{4}A_{R\mu\nu}^2-\frac{J^{\mu}_{\text{em}}}{\cos            \chi_0}A_{R\mu} -\frac{1}{4}S^2_{\mu\nu}\\ 
        &+\frac{1}{2}(A_{R\mu} S_{\mu})
        \left[ 
            \begin{array}{cc}
                m_\chi^2\sin^2\chi_0 & m_\chi^2\cos\chi_0\sin\chi_0 \\
                m_\chi^2\cos\chi_0\sin\chi_0 & m_\chi^2\sin^2\chi_0
            \end{array}
        \right]
        \left(
            \begin{array}{c}
                A_{R\mu} \\ S_{\mu}
            \end{array}
        \right),
        \label{eq:lagrangian_I}
    \end{aligned}
\end{equation}
where
\begin{equation}
    \begin{array}{ccc}
        A_R & =& \cos\chi_0 A \\
        S & =&X-\sin\chi_0 A.
    \end{array}
\end{equation}

In the Lagrangian $L_I$ (\ref{eq:lagrangian_I}), the photon and the additional boson are kinetically decoupled, with $A_{R\mu}$ interacting directly with the electromagnetic current. Consequently, $A_{R\mu}$ represents the physically observable photon field in the dark photon model, while $S$ corresponds to the sterile component, i.e., the dark photon field.

In vacuum ($J^{\mu}_{\text{em}}=0$), the mass matrix in $L_I$ is diagonalized via the unitary transformation $U_{I \to m}$:
\begin{equation}
    U_{I\to m} = \left[ \begin{array}{cc} \cos\chi_0 & -\sin\chi_0 \\ \sin\chi_0 & \cos\chi_0 \end{array} \right],
\label{eq:tra_matrix_Im}
\end{equation}
yielding the diagonalized Lagrangian:
\begin{equation}
    L_m = -\frac{1}{4}A_{1\mu\nu}^2-\frac{1}{4}A^2_{2\mu\nu}+\frac{1}{2}(A_{1}^\mu A_{2}^\mu)
    \left[ 
        \begin{array}{cc}
        0 & 0 \\
        0 & m_{\chi}^2
        \end{array}
    \right]
    \left(
        \begin{array}{c}
            A_{1\mu} \\ A_{2\mu}
        \end{array}
    \right).
    \label{eq:lagrangian_Lm}
\end{equation}
The kinetic terms for $A_1$ and $A_2$ are fully diagonal, with masses $0$ and $m_\chi$, respectively. Therefore, $A_1$ and $A_2$ are the mass eigenstates for vacuum propagation, related to the interaction states $(A_{R\mu},S)$ via the transformation matrix $U_{I\to m}$:
\begin{equation}
    \left( \begin{array}{c} A_{1\mu} \\ A_{2\mu} \end{array}\right) = U_{I\to m}\cdot \left( \begin{array}{c} A_{R\mu} \\ S_\mu \end{array} \right).
\end{equation}

In a medium, photon propagation is governed by the medium's properties. Classical electrodynamics attributes this to an effective photon mass $m_\gamma$, where the real component modifies the light field's phase (absorption is neglected here; $\text{Im}[\bar{m}_\gamma]=0$). The electromagnetic current in the Lagrangian is then:
\begin{equation}
    J^{\mu}_{\text{em}}=\frac{m_{\gamma}^2}{2}A_R^\mu.
\end{equation}
Substitution into $L_I$ enables diagonalization of the mass matrix through the field rotation $U_{I\to M}$:
\begin{equation}
    U_{I \to M} = \left( 
    \begin{array}{cc}
        \cos\chi_0 & -\frac{\sin\chi_0 m_\chi^2}{m_\chi^2-m_\gamma^2} \\
        \frac{\sin\chi_0 m_\chi^2}{m_\chi^2-m_\gamma^2} & \cos\chi_0
    \end{array} \right).
    \label{eq:tr_matix_IM}
\end{equation}
The diagonalized Lagrangian becomes:
\begin{align}
    L_M& = -\frac{1}{4}M_{1\mu\nu}^2-\frac{1}{4}M^2_{2\mu\nu}+\frac{1}{2}(M_{1}^\mu M_{2}^\mu) \cdot \\ \nonumber 
   &\left[ 
        \begin{array}{cc}
        m_\gamma^2+\frac{m_\chi^2m_\gamma^2\sin^2\chi_0}{m_\chi^2} & 0 \\
        0 & m_\gamma^2-\frac{m_\chi^2m_\gamma^2\sin^2\chi_0}{m_\chi^2}
        \end{array}
    \right]
    \left(
        \begin{array}{c}
            M_{1\mu} \\ M_{2\mu}
        \end{array}
    \right),
    \label{eq:lagrangian_LM}
\end{align}
where $M_1$ and $M_2$ are medium-dependent eigenstates with diagonal kinetic and mass terms. These eigenstates relate to the interaction basis $(A_{R\mu},S)$ via:
\begin{equation}
    \left( \begin{array}{c} M_{1\mu} \\ M_{2\mu} \end{array}\right) = U_{I\to M} \cdot \left( \begin{array}{c} A_{R\mu} \\ S_\mu \end{array}\right).
\end{equation}
We analyze photon propagation through the interstellar medium in gravitational fields and calculate the oscillation probabilities between mass eigenstates. The medium-dependent eigenstates $(M_1,M_2)$ and vacuum mass eigenstates $(A_1,A_2)$ are connected via the unitary rotation matrix:
\begin{equation}
    U_{M \to m}=U_{I \to m}U^{-1}_{I \to M},
\end{equation}
where $U_{I \to m}$ and $U_{I \to M}$ are defined in Equation (\ref{eq:tra_matrix_Im}) and Equation (\ref{eq:tr_matix_IM}).
For a system initially in the mass eigenstates $(A_1(0),A_2(0))$, its time evolution follows:
\begin{equation}
    \left( \begin{array}{c}
    A_1(t) \\ A_2(t)
    \end{array} \right) = U_{M\to m} e^{-iHt} U^{-1}_{M\to m} \left( \begin{array}{c}
        A_1(0) \\ A_2(0)
    \end{array} \right),
\end{equation}
with $H$ being the diagonalized Hamiltonian. Photons emitted in the $(1,0)$ state (pure $A_1$) oscillate to $A_2$ with probability: 
\begin{equation}
    P_{A_1 \to A_2} = \sin^2 2\chi_0\frac{m_\gamma^4}{|m_\chi^2-m_\gamma^2|^2}\sin^2 \left(\frac{(m_\chi^2-m_\gamma^2)t}{4E_\gamma}\right).
\end{equation}
For astrophysical (incoherent) sources, the oscillatory term $\sin^2(\cdot)$ averages to $1/2$. Defining the mixing parameter $\varepsilon = \sqrt{2}\sin\chi_0$, the effective probability becomes:
\begin{equation}
    P_{A_1\to A_2} = \varepsilon^2 \frac{m_\gamma^4}{|m_\chi^2-m_\gamma^2|^2}.
    \label{eq:osc_prob_mass}
\end{equation}
Equation (\ref{eq:osc_prob_mass}) provides the oscillation probability between photon and dark photon mass states in a medium under weak external gravitational fields. Crucially, this result diverges from the interaction states oscillation probability (Equation (\ref{eq:Interation_osc_proba})) derived for media. In the regime $m_\chi \ll m_\gamma$, where the dark photon mass is significantly smaller than the medium’s effective photon mass, the mass states oscillation probability remains unsuppressed. This advantage is expected to be applied in detection experiments.

\section{Example of Gravitational Lensing Quasar Scheme}
\label{section_3}

Under gravitational field effects, mass eigenstates $(A_1,A_2)$ propagating along spacetime geodesics experience differential trajectory shifts (geodesic deviations), inducing decoherence between the eigenstates. Remarkably, gravitational fields can mimic the decoherence effects traditionally attributed to absorptive media—such as reactor core plasma \citep{danilov2019constraints,park2017detecting} or solar plasma in helioscope experiments \citep{redondo2008helioscope,an2013dark}—but with a critical distinction: absorptive media suppress oscillations by decohering interaction eigenstates $(A_R,S)$, while gravitational fields decohere mass eigenstates $(A_1,A_2)$ through spacetime curvature gradients.

\begin{figure}
    \centering
    \includegraphics[width=1.0\linewidth]{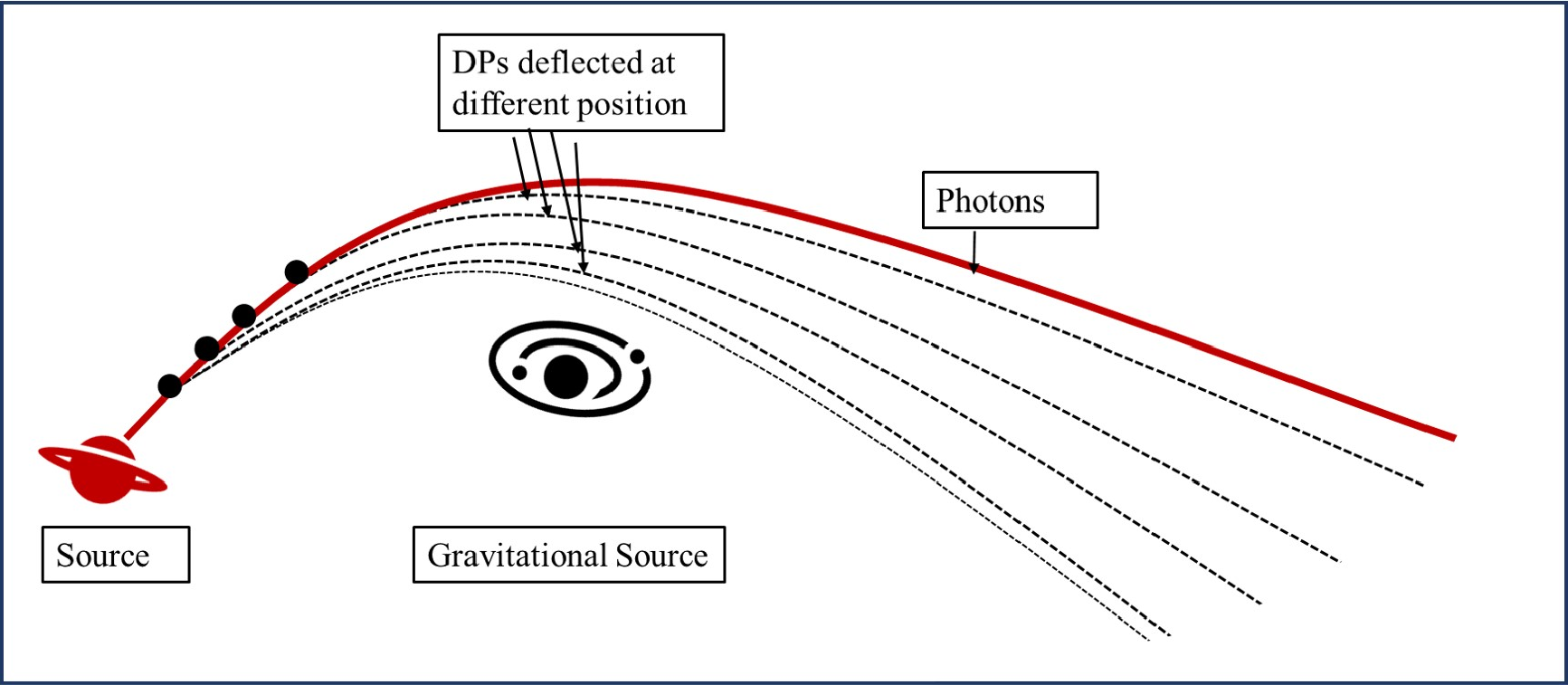}
    \caption{Illustration of this detection mechanism in a gravitationally lensed quasar system. Photons (red solid line) emitted from the quasar follow bent geodesics due to the foreground galaxy’s gravitational potential. The oscillated dark photon component (black dashed line) propagates along a distinct path, with the spatial separation between mass eigenstates suppressing quantum interference.}
    \label{fig:Illustration}
\end{figure}

Figure~\ref{fig:Illustration} schematically illustrates the astrophysical system under study. Photons emitted from a distant quasar are gravitationally deflected by a foreground galaxy and detected by observers.  In the dark photon framework, medium-induced oscillations between the mass eigenstates $(A_1,A_2)$ occur during propagation through the interstellar medium. Simultaneously, the foreground galaxy’s gravitational field induces geodesic separation between the eigenstates (black dashed line). This dual effect modifies the quasar’s apparent flux and introduces spectral distortions.

By analyzing flux discrepancies in gravitationally lensed quasars—compared to unlensed counterparts—we derive constraints on the dark-photon parameter space. In this section, we formulate the flux suppression caused by photon-dark photon oscillations in lensed systems, providing a quantitative tool for such analyses.

Consider a flux of $N_0$ photons propagating through the galactic medium. Decoherence between mass eigenstates arises over a characteristic distance $l$ due to gravitational geodesic deviation. The photon attenuation follows:
\begin{equation}
    dN = -N\frac{P}{l} dL,
    \label{eq:attenuation_diff}
\end{equation}
where $P$ is the oscillation probability between mass eigenstates, etc. Equation (\ref{eq:osc_prob_mass}). Integrating Equation (\ref{eq:attenuation_diff}) gives:
\begin{equation}
    N =N_0\exp{(-P\frac{L}{l})}=N_0\exp{(-P\tau)},
    \label{eq:N}
\end{equation}
where $\tau \equiv L/l$ is the optical depth quantifying the effective number of decoherence events. The attenuation factor $N/N_0$ corresponds to the observed quasar luminosity suppression.

To estimate $\tau$, we compute the differential angular deflection $\Delta\alpha$ between photons and dark photons using weak-field general relativity:
\begin{equation}
    \Delta \alpha = \frac{2GM}{R_{gal}c^2}\left( \frac{m_\chi^2}{E^2-m_\chi^2} \right).
\label{eq:angle_diff}
\end{equation}
Here, $R_{gal}$ is the gravitational lens’s characteristic scale, and the full derivation is given in Appendix \ref{section:appendix}. 

The maximum path-length difference between photons and dark photons, induced by their differential deflection, is approximated as:
\begin{equation}
    L\approx L_{\text{s}}\times \Delta \alpha,
\end{equation}
where $L_s$ is the quasar-Earth distance. For a photon energy $E$, the coherence length of the thermal photon wavepacket is\citep{donges1998coherence}:
\begin{equation}
    \Delta h = \frac{197MeV \cdot fm}{E},
\end{equation}
which corresponds to the de Broglie wavelength scale. While the actual propagation paths include non-geodesic corrections, the sub-eV dark photon mass ensures minimal trajectory separation, justifying a linearized approximation for the optical depth:
\begin{equation}
    \tau = \frac{L}{\Delta h}.
    \label{eq:tau}
\end{equation}
Substituting Equations (\ref{eq:angle_diff})-(\ref{eq:tau}) into Equation (\ref{eq:N}), the quasar’s flux suppression becomes:
\begin{equation}
    \frac{N}{N_0} \approx 1-\varepsilon^2\frac{m_\gamma^4}{|m_\chi^2-m_\gamma^2|^2} \frac{L_{\text{s}}E}{197MeV \cdot fm}\frac{2GM}{R_{gal}c^2}\left( \frac{m_\chi^2}{E_\gamma^2-m_\chi^2} \right).
    \label{eq:N/N_0}
\end{equation}
The $1/(E_\gamma^2-m_\chi^2)$ dependence in Equation (\ref{eq:N/N_0}) indicates stronger attenuation at lower photon energies $(E_\gamma \gg  m_\chi)$, leading to spectral hardening.

\section{Results and Discussions}
\label{section_4}

The quasar luminosity function quantifies the number density of quasars as a function of intrinsic brightness and redshift \citep{gaston1983luminosity,notini1972luminosity,pei1995luminosity}. By analyzing redshift-binned luminosity distributions, cosmological evolution of quasar populations can be reconstructed\citep{hawkins1995evolution}. We extend this framework to gravitationally lensed quasars, comparing their luminosity function with unlensed populations. Gravitational lensing-induced chromatic dispersion—arising from photon-dark photon oscillations—suppresses the observed flux of lensed quasars.

A statistically significant signature emerges when the flux attenuation exceeds an assumed threshold of $ 10\% $ ($N/N_0<0.1$), which we conservatively adopt based on instrumental sensitivity limits and astrophysical background uncertainties in current surveys.

We adopt typical parameters for quasar gravitational lensing systems: a source redshift $z=1$ (comoving distance $L_s$), a foreground galaxy radius $R_{gal} = 1kpc$, and a galaxy mass $M=10^{12}M_\odot$. The analysis assumes optical wavelength observations ($E_\gamma\sim eV$). The effective photon mass $m_\gamma$ in the interstellar medium scales with cosmological evolution; for low-redshift regions (z<1), $m_\gamma\sim 10^{-14}eV$\citep{mirizzi2009microwave}, consistent with the plasma frequency of the interstellar medium.

\begin{figure}

\includegraphics[width=1.0\linewidth]{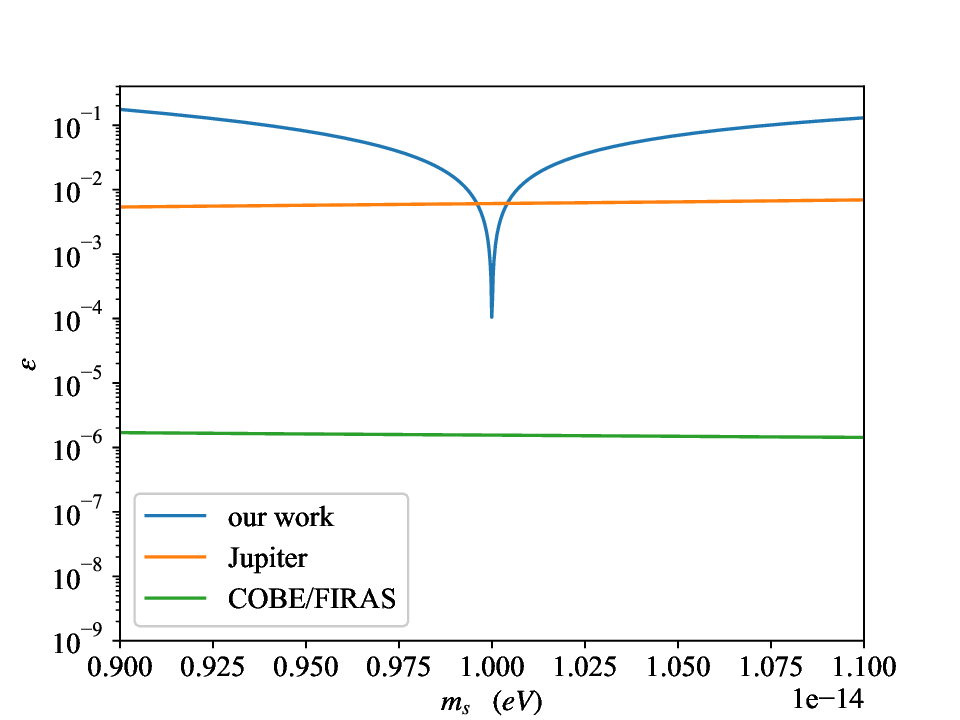}
\caption{The sensitivity of this quasar lensing-based dark photon detection scheme. The blue solid line represents our method’s sensitivity, which complements existing constraints: the orange solid line shows limits from the Jupiter magnetic field experiment \citep{marocco2021dark,yan2024constraints}, and the green solid line indicates COBE/FIRAS \citep{fixsen1996cosmic,mirizzi2009microwave,mcdermott2020cosmological,caputo2020dark,garcia2020effective} bounds on CMB spectral distortions. 
\label{fig:result}}
\end{figure}

As shown in Figure~\ref{fig:result}, this method achieves a sensitivity of $\sim 10^{-2}$ for dark photon masses in the range $(0.975\times 10^{-14}eV,1.025\times 10^{-14}eV)$. A striking enhancement to $\sim 10^{-4}$ occurs at $m_\gamma \approx 10^{-14}eV$, driven by resonant photon-dark photon oscillations when $m_\chi \sim m_\gamma$, as evident from the $1/|m_\chi^2-m_\gamma^2|^2$  dependence in Equation (\ref{eq:osc_prob_mass}).

Our analysis assumes a uniform interstellar medium density, corresponding to a fixed effective photon mass $m_\gamma = 10^{-14}eV$, which produces a sharp resonance peak. In realistic astrophysical environments, however, density variations in the interstellar medium would broaden the resonance into a smoother distribution, reflecting the range of $m_\gamma$ values encountered along the line of sight.

Relative to the Jupiter magnetic experiment \citep{marocco2021dark,yan2024constraints} and COBE/FIRAS \citep{fixsen1996cosmic,mirizzi2009microwave,mcdermott2020cosmological,caputo2020dark,garcia2020effective}, our method do not achieve sensitivity improved. Future advancements in data analysis algorithms or observational precision could enhance sensitivity. Additionally, standardizable Type Ia supernovae—statistically robust distance indicators—could provide complementary constraints, leveraging their well-calibrated luminosity profiles.

While the gravitational lensing scheme cannot fully exploit the mass-independent nature of photon-dark photon oscillations in gravitational fields, we propose that neutron star evolution and black hole accretion disk systems may exhibit more pronounced signatures of gravitationally induced dark photon decoherence. These systems, characterized by extreme gravitational fields and high-energy environments, offer unique opportunities to probe dark photon interactions. However, their complex dynamics and emission mechanisms require detailed modeling to disentangle dark photon effects from astrophysical backgrounds.

\section{Conclusion and prospect}
\label{section_5}

We propose a novel oscillation mechanism between photon and dark photon mass eigenstates during propagation in media under gravitational fields, deriving the oscillation probability formula. Unlike conventional dark photon detection schemes, this mechanism is immune to medium-induced suppression of oscillations. Furthermore, we suggest that gravity can induce photon–dark-photon mass-state separation during their propagation through interstellar media. By observing the potential attenuation in gravitationally lensed quasar luminosities, we seek to constrain the dark-photon parameter space. Under typical system-scale assumptions, our preliminary estimate indicates that for dark-photon masses in the range $(0.975\times 10^{-14}eV,1.025\times 10^{-14}eV)$, the sensitivity to the mixing parameter $\varepsilon$ can reach approximately $10^{-2}$. This result can be further improved with additional data and a more detailed analysis. Finally, we anticipate that this oscillation mode may be fruitfully exploited in neutron-star and black-hole accretion-disk systems.


\bibliography{refs}

\appendix
\section{Deflection Angle between Photon and Dark-Photon}
\label{section:appendix}
\setcounter{equation}{0}
\setcounter{figure}{0}

\subsection[\appendixname~\thesection]{Calculation basis}
This appendix presents calculation process of gravitational lensing deflection angle of massive particle and massless particle. Firstly, gravity of foreground is described by the Schwarzschild metric with isotropic static assuming.
\begin{align}
    ds^2=&g_{\mu\nu}dx^\mu dx^\nu=-(1-\frac{2M}{r})dt^2+(1-\frac{2M}{r})^{-1}dr^2 \nonumber \\
         &+r^2d\theta^2+r^2\sin^2\theta d\phi^2.
\end{align}
The Lagrangian can be written as
\begin{eqnarray}
    L_G=-\frac{1}{2}g_{\mu\nu}\dot x^\mu \dot x^\nu 
    &=&\frac{1}{2}(1-\frac{2M}{r})(\frac{dt}{d\lambda})^2+(1-\frac{2M}{r})^{-1}(\frac{dr}{d\lambda})^2 \nonumber \\
    & &+r^2(\frac{d\theta}{d\lambda})^2+r^2\sin^2\theta (\frac{d\phi}{d\lambda})^2,
    \label{Eq:Lg}
\end{eqnarray}
where $\lambda$ is affine parameter which could be defined as proper time for massive particle. The Lagrangian bases on four-velocity of particle, thus
\begin{equation}
    L_G=\kappa=\left\{ \begin{matrix}
        0\quad (massless \quad particle)\\
        \frac{1}{2} \quad (massive\quad particle)
    \end{matrix}\right. .
    \label{Eq:kappa}
\end{equation}
The isotropic and static condition suggest two first integrals
\begin{equation}
    \frac{\partial L_G}{\partial t}=0,\qquad \frac{\partial L_G}{\partial \phi}=0.
\end{equation}
It means two conserved quantities, energy and angular momentum.
\begin{subequations}
    \begin{align}
        E=&\frac{\partial L_G}{\partial \dot t}=(1-\frac{2M}{r})\frac{dt}{d\lambda}\label{AAEq:E},\\
        L=&\frac{\partial L_G}{\partial \dot \phi}=r^2\sin^2 \frac{d\phi}{d\lambda}\label{AAEq:L}.
    \end{align}
\label{AAEq:E_L}
\end{subequations}
Combining Equation (\ref{Eq:Lg}) and Equation (\ref{Eq:kappa}), we obtain normalization condition of four-velocity.
\begin{equation}
    \left(\frac{dr}{d\lambda}\right)^2=E^2-(1-\frac{2M}{r})(2\kappa-\frac{L^2}{r^2}).
    \label{AAEq:4V}
\end{equation}
The calculation base on  Equation (\ref{AAEq:E_L}) and Equation (\ref{AAEq:4V}) and would be divided into massive section and massless section.

\subsection[\appendixname~\thesection]{Deflection of massless particle(photon)}
This section discusses the massless particle deflection in gravity.
Set $\kappa=0$, reduce $\lambda$ and get orbital equation by combining Equation (\ref{AAEq:E_L}) and Equation (\ref{AAEq:4V}).
\begin{equation}
    \left[\frac{d}{d\phi}(\frac{1}{r})\right]^2=\left[\frac{E}{L}\right]^2-\frac{1}{r^2}(1-\frac{2M}{r}).
    \label{AAEq:orbital_eq}
\end{equation}
Equation (\ref{AAEq:orbital_eq}) derivation of $\phi$ can be simplified as
\begin{equation}
    \frac{d^2u}{d\phi^2}+u=3Mu^2,
    \label{AAEq:orbital_eq_more}
\end{equation}
where $u=1/r$.Equation (\ref{AAEq:orbital_eq_more}) and original conditions contains orbit information of massless particle. The last work is solving the differential equation. The nonlinear feature makes it hard. The nonlinear term is contributed by gravity. For gravity is small in gravitational lensing system, the term of $3Mu^2$ is regard as perturbation. The solution of Equation (\ref{AAEq:orbital_eq_more}) is

\begin{figure}
   \centering
\includegraphics[width=1.0\linewidth]{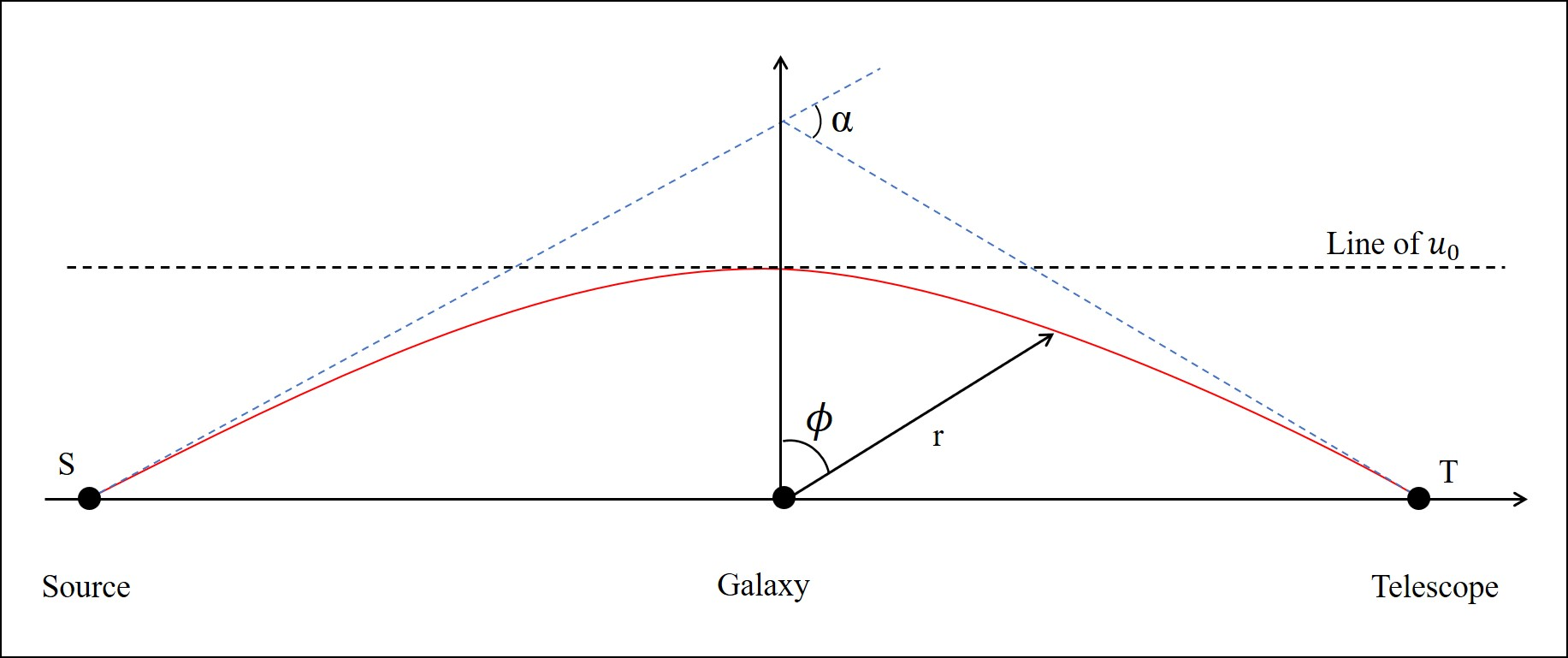}
\caption{Illustration of parameters of massless and massive particle geodesic.} 
\label{fig:appendix_lensing_Illustration}
\end{figure}

\begin{equation}
    u = u_0+\xi u_1 +o(\xi^2),\qquad where\quad \xi=3M.
    \label{AAEq:u_perburbation}
\end{equation}
Substituting Equation (\ref{AAEq:u_perburbation}) into Equation (\ref{AAEq:orbital_eq_more}), we have
\begin{subequations}
    \begin{align}
        \frac{d^2u_0}{d\chi^2}+u_0=0, \label{AAEq:EOM_perturbation_1a}\\
        \frac{d^2u_1}{d\chi^2}+u_1=u_0^2.\label{AAEq:EOM_perturbation_2a}
    \end{align}
\end{subequations}
Equation (\ref{AAEq:EOM_perturbation_1a}) describes a scene without gravity, so the solution is
\begin{equation}
    u_0 = \frac{\cos\phi}{R_{galaxy}}.
    \label{AAEq:u_0}
\end{equation}
Hence, the first order approximate solution of Equation (\ref{AAEq:orbital_eq_more}) is
\begin{equation}
    u = \frac{\cos\phi}{R_{galaxy}}+\frac{M}{R_{galaxy}^2}(1+\sin^2\phi).
    \label{AAEq:u}
\end{equation}
Limiting r to infinity,$ u\to 0$, Equation (\ref{AAEq:u_0}) suggests $\phi'=\pm \pi/2$. Deflection angle in the first order approximation Equation (\ref{AAEq:u}) can be defined as
\begin{equation}
    \phi = \phi '\pm \alpha'=\pm (\frac{\pi}{2}+\alpha ' ).
    \label{AAEq:phi}
\end{equation}
$\alpha'$ is a half of total deflection angle. Equation (\ref{AAEq:u}) and Equation (\ref{AAEq:phi}) derive the deflection angle of photon in gravitational field.
\begin{equation}
    \alpha_{massless} =2\alpha '= \frac{4M}{R_{galaxy}}=\frac{4GM}{R_{galaxy}c^2}.
    \label{eq:massless_ref_angle}
\end{equation}

\subsection[\appendixname~\thesection]{Deflection of massive particle}
In this section, we focus on massive particle orbital character in gravity.
Set $\kappa = 1/2$, deal like Equation (\ref{AAEq:orbital_eq}) and Equation (\ref{AAEq:orbital_eq_more}), and we have
\begin{equation}
    \frac{d^2u}{d\phi^2}+u=\frac{M}{L^2}+3Mu^2.
    \label{AAEq:massive_eom}
\end{equation}
We concern the low mass range of DP in order to obtain lower undetectable mass limit about our work. $M/L^2$ is bigger than $3Mu^2$ in classical physics. But for small mass condition, the state is reversed. $M/L^2$ need be regarded as first order approximation at least. Using the same perturbation strategy, the solution Equation (\ref{AAEq:u_0}) is also the zeroth order approximation solution of Equation (\ref{AAEq:massive_eom}). Taking Equation (\ref{AAEq:u_0}) into Equation (\ref{AAEq:massive_eom}) gives
\begin{equation}
    \frac{d^2u_1}{d\phi^2}+u_1=\frac{M}{L^2}+3M\left({\frac{\cos\phi}{R_{galaxy}}}\right)^2.
\end{equation}
Its solution is under the first order approximation showing as
\begin{equation}
    u_1=\frac{M}{L^2}+\frac{M}{R_{galaxy}^2}(1+\sin^2\phi).
\end{equation}
So, the solution of Equation (\ref{AAEq:massive_eom}) in the first order approximation is
\begin{equation}
    u = \frac{\cos\phi}{R_{galaxy}}+\frac{M}{L^2}+\frac{M}{R_{galaxy}^2}(1+\sin^2\phi).
\end{equation}
Deflection angle $\alpha''$ can be defined in
\begin{equation}
    \phi = \pm(\frac{\pi}{2}+\alpha '').
    \label{AAEq:total_angle}
\end{equation}
We could obtain the total deflection angle by combining with Equation (\ref{AAEq:total_angle}) and $u=0$.
\begin{equation}
    \alpha_{massive} = 2\alpha ''=\frac{2MR_{galaxy}}{L^2}+\frac{4M}{R_{galaxy}}.
\end{equation}
$L$ corresponds to angle momentum of per rest mass,
\begin{equation}
    L = \frac{\vec{r}\times \vec{p}}{m_0}=\frac{R_{galaxy}m\beta}{m_0}=\frac{\beta}{\sqrt{1-\beta^2}}R_{galaxy},
\end{equation}
where $\beta=v/c$.
Therefore,
\begin{equation}
    \alpha_{massive} = \frac{2M}{R_{galaxy}\beta^2}+\frac{2M}{R_{galaxy}}= \frac{2GM}{R_{galaxy}v^2}+\frac{2GM}{R_{galaxy}c^2}.
    \label{eq:massive_ref_angle}
\end{equation}
\subsection[\appendixname~\thesection]{the differential angular deflection }
The differential angular deflection can be derived by taking the difference between Equation (\ref{eq:massive_ref_angle}) and Equation (\ref{eq:massless_ref_angle}):
\begin{equation}
    \Delta \alpha = \frac{2GM}{R_{galaxy}c^2}\left( \frac{m_\chi^2}{E^2-m_\chi^2} \right).
\end{equation}

\end{document}